\documentclass[twocolumn,secnumarabic,amssymb,nobibnotes, aps, prl]{revtex4-2}
\usepackage{graphicx}
\usepackage{dcolumn}
\usepackage{bm}
\usepackage[breaklinks]{hyperref}
\hypersetup{
    colorlinks=true,
    linkcolor=blue,
    citecolor=blue
}

\usepackage[version=4]{mhchem}
\usepackage{natbib}
\usepackage{url}
\usepackage{textcase}
\usepackage{xcolor}
\usepackage{braket}
\usepackage{amsmath}
\usepackage[normalem]{ulem}
\begin{document}

\title{Magnon-phonon interactions from first principles}

\author{Khoa B. Le$^{1}$, Ali Esquembre-Ku\v{c}ukali\'c$^{2}$, Hsiao-Yi Chen$^{3}$, Ivan Maliyov$^{1}$, Yao Luo$^{1}$, \\ Jin-Jian Zhou$^{4}$, Davide Sangalli$^{5}$, Alejandro Molina-S\'anchez$^{2}$, Marco Bernardi$^{1}$}

\affiliation{$^{1}$
 Department of Applied Physics and Materials Science, and Department of Physics,\\ California Institute of Technology, Pasadena, California 91125, USA
}%
\affiliation{$^{2}$
 Institute of Materials Science (ICMUV), University of Valencia,  Catedr\'{a}tico Beltr\'{a}n 2,  E-46980,  Valencia,  Spain
}%
\affiliation{$^{3}$
 Institute for Materials Research, Tohoku University, Sendai, 980-8577, Japan
}%

\affiliation{$^{4}$
 School of Physics, Beijing Institute of Technology, Beijing 100081, China
}%
\affiliation{$^{5}$
 Istituto di Struttura della Materia-CNR (ISM-CNR), 
 Area della Ricerca di Roma 1, Monterotondo Scalo, Italy
 \vspace{14pt}
}

\begin{abstract}
Modeling spin-wave (magnon) dynamics in novel materials is important to advance spintronics and spin-based quantum technologies. The interactions between magnons and lattice vibrations (phonons) limit the length scale for magnon transport. However, quantifying these interactions remains challenging. 
Here we show many-body calculations of magnon-phonon (mag-ph) coupling based on the \textit{ab initio} Bethe-Salpeter equation. We derive expressions for mag-ph coupling matrices and compute them in 2D ferromagnets, focusing on hydrogenated graphene and monolayer CrI$_3$. Our analysis shows that electron-phonon ($e$-ph) and mag-ph interactions differ significantly, where modes with weak $e$-ph coupling can exhibit strong mag-ph coupling (and vice versa), and reveals which phonon modes couple more strongly with magnons.
In both materials studied here, the inelastic magnon relaxation times decrease abruptly above the threshold for emission of strongly coupled phonons, revealing a low-energy window where magnons are long-lived. Averaging over this window, we compute the temperature-dependent magnon mean-free path, a key figure of merit for spintronics, entirely from first principles.
The theory and computational tools shown in this work enable studies of magnon interactions, scattering, and dynamics in generic materials, advancing the design of magnetic systems and magnon- and spin-based devices.
\end{abstract}

\maketitle
\textit{Introduction.\textemdash} Spin waves, or magnons, are collective spin excitations in magnetic materials. The availability of novel magnetic quantum materials~\cite{basov2017towards, kagome_ye2018}, and the discovery of magnetism in layered van der Waals crystals such as \ce{CrX3} (X \!=\! Cl, I, Br) and others~\cite{Lee2016, Gong2017, Huang2017, Lee2021}, provide a wide array of platforms for magnon physics~\cite{Li2022, Roche2024}.  Magnons play a central role in emerging quantum technologies as they can be manipulated with electric fields, converted to free spins~\cite{Zhang2012_Aug,Bender2012}, interfaced with photons or qubits in hybrid quantum magnonic setups~\cite{Yuan2022}, and used as information carriers for computing and data \mbox{storage~\cite{Bader2010, Kruglyak2010, Chumak2015, Flebus2024}.} 
\\
\indent
Achieving long magnon coherence times is essential to advance these magnon-based applications. Magnon decoherence can result from several effects, including magnon-magnon interactions~\cite{Lindner2003}, coupling of magnons with conduction electrons \cite{Muller2019, Ritzmann2020, Nabok2021, Beens2022}, and magnon-phonon (mag-ph) interactions~\cite{Maehrlein2018, Hioki2022}.
Magnon-phonon coupling has received particular attention: recent work has studied the formation of strongly coupled mag-ph states \cite{Godejohann2020}, coherent oscillations between magnons and phonons~\cite{Hioki2022}, hybrid mag-ph crystals~\cite{hybrid_magph}, tuning of mag-ph interactions with applied magnetic fields \cite{Berk2019}, and the role of mag-ph coupling in spin-based thermal and thermoelectric effects \cite{Boona2014, Kikkawa2016}. Given their broad relevance, developing accurate calculations of mag-ph interactions is important to advance magnon physics. 
\\
\indent
A widely used framework for magnon calculations is the Heisenberg spin model, often with the inclusion of anisotropic and antisymmetric exchange terms~\cite{Dzyaloshinsky1958, Moriya1960, Toth2015}. Time-dependent DFT has also been used to study magnons and more recently mag-ph coupling \cite{Delugas2023, Gorni2023}. In the Heisenberg model, the exchange parameters are typically extracted from inelastic neutron scattering experiments or computed from first principles using density functional theory (DFT)~\cite{ xiang2012magnetic, lkag, tb2j}. 
The mag-ph coupling can be studied by analyzing the effects of lattice perturbations. This approach has been used to study mag-ph interactions and relaxation times \cite{Wang2020, Cong2022}, mag-ph hybridization \cite{Cui2022, Metzger2024}, and coupled mag-ph transport \cite{Flebus2017}.
Although they are efficient to compute, mag-ph interactions from the Heisenberg model depend on the choice of exchange parameters and are tied to the phenomenological form of the Hamiltonian, which may fail to capture the complexity of real materials.  
Developing quantitative approaches to compute magnon dispersions and interactions on the same footing would deepen the understanding of mag-ph coupling in materials. 
\\  
\indent
In this Letter, we show a formalism for studying mag-ph interactions using the \textit{ab initio} Bethe-Salpeter equation (BSE) and develop corresponding first-principles calculations.
This framework provides a microscopic understanding of mag-ph interactions, enabling predictions of magnon relaxation times and mean-free paths, discovery of phonon modes with strong coupling to magnons, and momentum-space analysis of mag-ph coupling. In two paradigmatic systems, hydrogenated monolayer graphene (H-MLG) and monolayer CrI$_3$, our approach reveals striking differences between $e$-ph interactions, routinely studied from first principles, and the mag-ph interactions computed here. Our approach paves the way for quantitative studies of the diverse physics of coupled phonons and magnons, with broad applications in spintronics and magnon-based devices. 

\begin{figure}[!t] %
    \includegraphics[width=\linewidth]{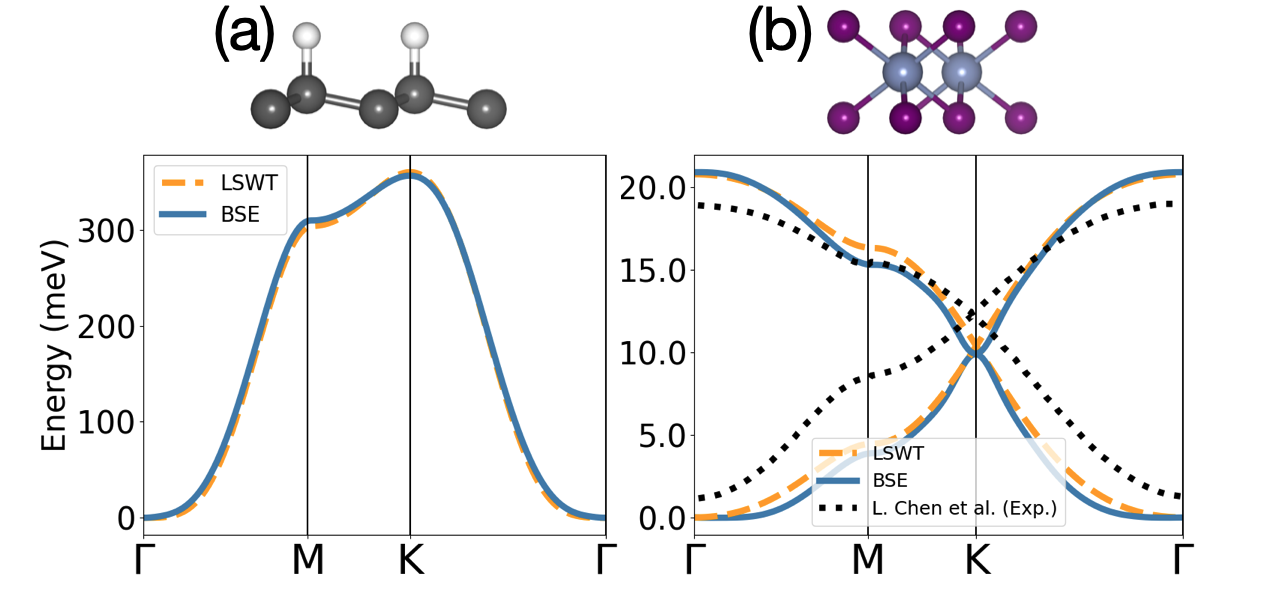}
    \caption{\label{fig:lattice-disp}Magnon dispersions in (a) H-MLG and (b) monolayer \ce{CrI3}, computed with the finite-momentum magnon BSE. Results from the Heisenberg model with up to third nearest-neighbor exchange couplings and experimental data from Ref.~\cite{Lebing2018} are shown for comparison.}
\end{figure}

\textit{Magnon dispersions.\textemdash} The BSE formalism is based on two-particle Green's functions and is widely used to study excitons~\cite{Onida2002, Rohlfing2000, Palummo2015} and more recently exciton-phonon interactions~\cite{Chen2020, Chen2022, exph_antonius, exph_offdiag}.
Studies of magnetic excitations using the BSE are a recent direction, with only a few examples reported so far~\cite{Olsen2021}. 
In the BSE, a magnon is regarded as a correlated pair of electron and hole with opposite spins, analogous to a spin-flip exciton. The effective BSE for magnons with finite momentum $\mathbf{Q}$ is written as
\begin{equation}
\sum_{v'c'} \mathcal{H}^{\rm mag}_{vc,v'c'}(\mathbf{Q}) \Psi^{\lambda_{n}}_{v'c'}  = E_{\lambda_{n}}(\mathbf{Q}) \Psi^{\lambda_{n}}_{v'c'}~\vspace{-10pt},
\end{equation}
where the exchange interaction is absent in the magnon BSE Hamiltonian, $\mathcal{H}^{\rm mag}$, due to the opposite electron-hole spin~\cite{Marsili2021}. Solving the magnon BSE provides the energies $E_{\lambda_{n}}(\mathbf{Q})$, for magnons with mode index $\mathit{n}$ and center-of-mass momentum $\mathbf{Q}$, and the corresponding wave functions expressed in the electron-hole transition basis~\cite{Gatti2013},  $\ket{\lambda_{n}}=\sum_{vc}\Psi^{\lambda_{n}}_{vc}\ket{v \uparrow, c \downarrow}$.
Our companion paper~\cite{companion} provides a more extensive analysis of the BSE for magnons, including distinguishing exciton and magnon solutions, addressing the Goldstone sum rule violation, and inclusion of spin-orbit coupling (SOC) effects.
\\
\indent  
We first compute magnon dispersions in freestanding H-MLG and monolayer CrI$_3$. We calculate their electronic structure using spin-polarized plane-wave DFT calculations (without SOC) with the \textsc{Quantum ESPRESSO} package~\cite{Giannozzi2009}, and solve the magnon BSE as implemented in the \textsc{YAMBO} code \cite{Sangalli2019, Marsili2021}, using a scissor correction to obtain band gaps that match previous GW calculations~\cite{molina2020magneto} (see below for details~\cite{Comp.detail}).

\begin{figure}[htb]
\includegraphics[width=1.0\linewidth]{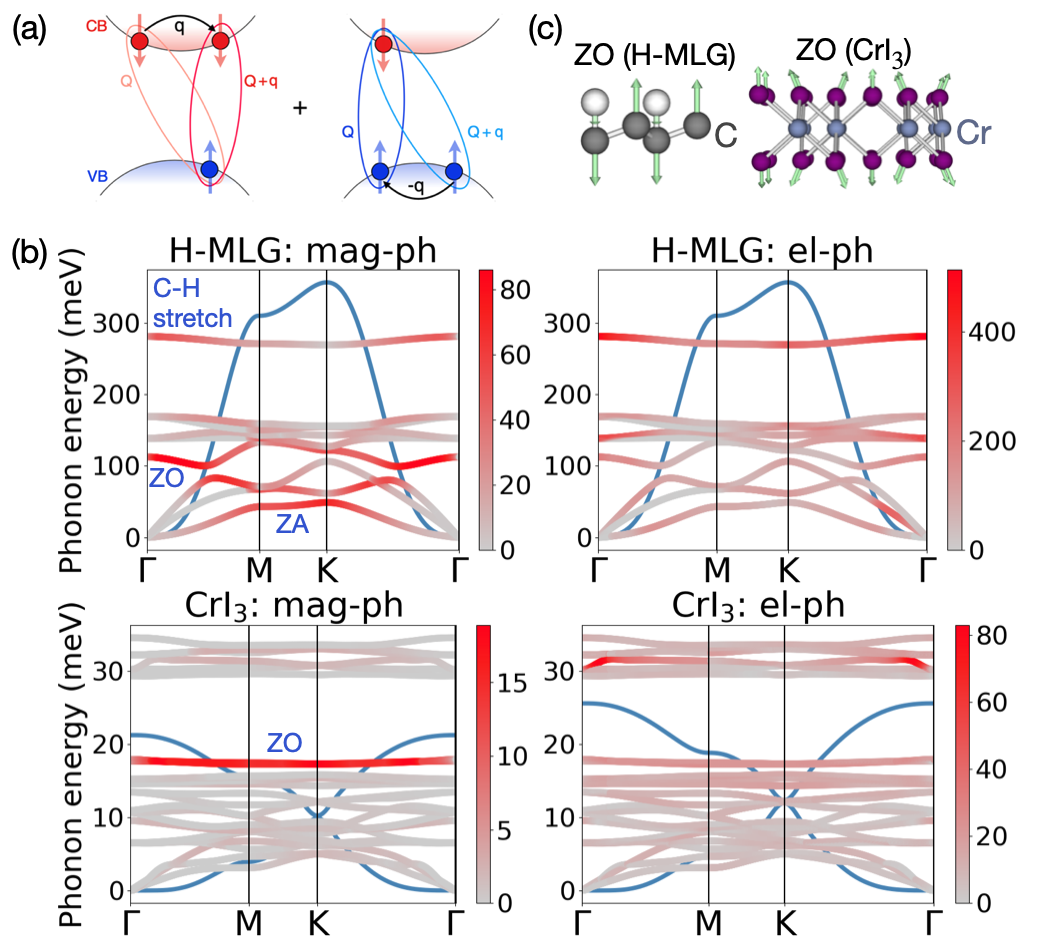}
\caption{\label{fig:magph-mat}(a) The mag-ph interaction viewed as a superposition of electron- (red) and hole-phonon (blue) scattering processes between initial and final magnon states, shown as ovals. (b) Phonon dispersion in H-MLG and \ce{CrI3} overlaid with a color map of average mag-ph and $e$-ph coupling strengths for each phonon mode, $|\mathcal{G}_{\nu}(\mathbf{q})|$ and $|g_{\nu}(\mathbf{q})|$, respectively. The mag-ph coupling in CrI$_3$ is an average of acoustic and optical magnons at $\mathbf{Q}=0$. The blue lines show the magnon dispersions in both materials.
(c) Phonon modes with the largest mag-ph coupling in H-MLG and \ce{CrI3}.}
\end{figure}

In Figure~\ref{fig:lattice-disp}, we show a comparison of the magnon dispersions obtained from the BSE and the Heisenberg model. The Heisenberg model is solved within linear spin wave theory (LSWT), as implemented in the \textsc{SpinW} code~\cite{Toth2015}, with exchange interactions up to third nearest neighbors obtained by fitting the BSE results. The convergence of the BSE calculations, the effect of SOC, and the exchange coupling constants used in LSWT are discussed in the companion paper~\cite{companion}. 
In particular, SOC opens a small gap ($0.3-1.8$ meV) between the acoustic and optical magnons at $\mathbf{Q}=K$~\cite{Olsen2021, companion}, and our BSE calculations with SOC can accurately predict this gap~\cite{companion}.
\\
\indent
\textit{Magnon-phonon coupling.\textemdash} Magnons and excitons are normally viewed as distinct quasiparticles, but in the framework of the BSE, they are both described as coherent superpositions of electron-hole transitions $–$ in the case of magnons, spin-flip transitions in a magnetic material~\cite{Marsili2021, Olsen2021}. This shared structure forms the basis for studying mag-ph interactions using the BSE. We write the Hamiltonian for coupled magnons and phonons as
\begin{align}
\label{eq:H_mag-ph}
&\mathcal{H}_{\text{mag-ph}}=\sum_{n\mathbf{Q}} E_{\lambda_{n}}(\mathbf{Q})\hat{S}^{\dagger}_{n\mathbf{Q}}\hat{S}_{n\mathbf{Q}} + \sum_{\nu\mathbf{q}} \hbar\omega_{\nu\mathbf{q}}\hat{b}^{\dagger}_{\nu\mathbf{q}}\hat{b}_{\nu\mathbf{q}} \nonumber \\
&\quad+\sum_{\substack{nm\mathbf{Q} \\ \nu \mathbf{q}}}
\mathcal{G}_{nm\nu}(\mathbf{Q},\mathbf{q})\hat{S}^{\dagger}_{m (\mathbf{Q+q})}\hat{S}_{n\mathbf{Q}}(\hat{b}_{\nu\mathbf{q}}+\hat{b}^{\dagger}_{\nu\mathbf{-q}}),
\end{align}
where we use creation and annihilation operators for magnons ($\hat{a}^{\dagger}$ and $\hat{a}$) and phonons ($\hat{b}^{\dagger}$ and $\hat{b}$) and introduce the mag-ph coupling matrices, $\mathcal{G}_{nm\nu}(\mathbf{Q},\mathbf{q})$, quantifying the probability amplitude for scattering from an initial magnon state $\ket{\lambda_{n}}$ with momentum $\mathbf{Q}$ to a final state $\ket{\lambda_{m}}$ with momentum $\mathbf{Q} + \mathbf{q}$, by emitting or absorbing a phonon with mode index $\nu$ and wave vector $\mathbf{q}$. The bilinear mag-ph mixing term, left out in the expression above, can also be computed in the BSE framework, as we show in the Supplemental Material (SM)~\cite{SM}. The mag-ph mixing is proportional to the SOC strength, which is weak in the materials studied here, where it has a negligible effect on the magnon dispersions, as we verify with explicit calculations~\cite{SM}.
\\
\indent
Following the BSE formalism for exciton-phonon interactions~\cite{Chen2020}, we treat atomic displacements as first-order perturbations to the BSE magnon wave function, and write the mag-ph coupling as a superposition of e-ph scattering processes weighted by the spin-flip transition amplitudes contributing to the magnon state [see Fig.~\ref{fig:magph-mat}(a)] (see SM~\cite{SM} for the derivation):
\begin{widetext}
\begin{align}
\label{eq:mag-ph}
\mathcal{G}_{nm\nu}(\mathbf{Q},\mathbf{q})= \sum_{\mathbf{k}}
\Big[ 
\sum_{vcc{}'}\Psi^{\lambda_{m}(\mathbf{Q}+\mathbf{q})*}_{\substack{v\mathbf{k},\uparrow \\ c(\mathbf{k}+\mathbf{Q}+\mathbf{q}),\downarrow}}\Psi^{\lambda_{n}(\mathbf{Q})}_{\substack{v\mathbf{k},\uparrow \\ c{}'(\mathbf{k}+\mathbf{Q}),\downarrow}} g_{cc{}'\nu,\downarrow}(\mathbf{k}+\mathbf{Q},\mathbf{q})
-\sum_{vcv{}'} \Psi^{\lambda_{m}(\mathbf{Q}+\mathbf{q})*}_{\substack{v(\mathbf{k}-\mathbf{q}),\uparrow \\ c(\mathbf{k}+\mathbf{Q}),\downarrow}}\Psi^{\lambda_{n}(\mathbf{Q})}_{\substack{v{}'\mathbf{k},\uparrow \\ c(\mathbf{k}+\mathbf{Q}),\downarrow}} g_{v{}'v\nu, \uparrow}(\mathbf{k}-\mathbf{q},\mathbf{q})
\Big]
\end{align}
\end{widetext}

\noindent 
where the $e$-ph matrix elements $g_{vv'\nu, \uparrow (\downarrow)}(\mathbf{k},\mathbf{q})$, \mbox{defined} as in Refs.~\cite{Bernardi2016,Zhou2021}, are amplitudes of phonon-mediated scattering for valence electronic states, and $g_{cc'\nu, \uparrow (\downarrow)}(\mathbf{k},\mathbf{q})$ for conduction states, where the spin of the electron (up or down) is conserved in the absence of SOC. Physically, our magnon-phonon coupling captures the effect of atomic displacements (phonons) on the electronic states entering the magnon wave function, which alters the structure and energy of magnon excitations. This formalism differs from the traditional picture of atomic motions modulating exchange parameters, offering a microscopic, wave function-based treatment.
\begin{figure}[t!]
\centering
\includegraphics[width=\linewidth]{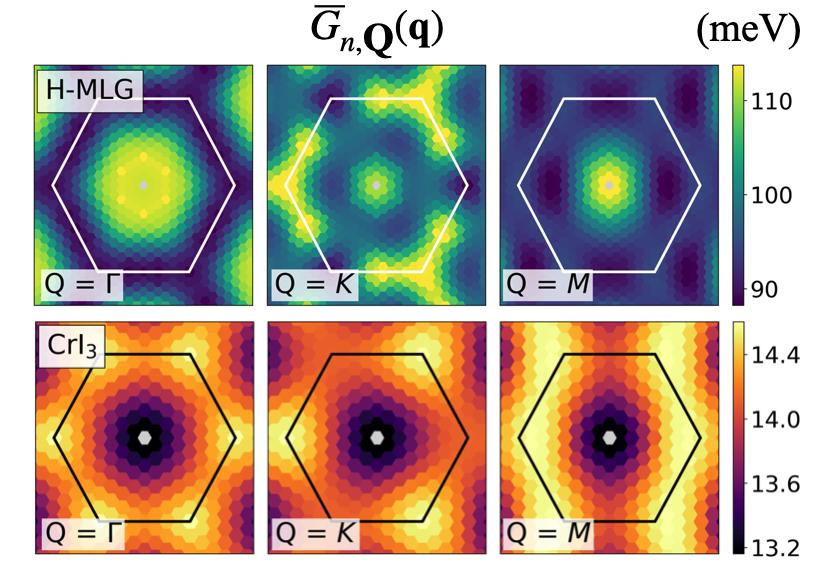}
\caption{\label{fig:magph-vs-q}  Brillouin-zone maps of average mag-ph coupling strengths, $\overline{G}_{n\textbf{Q}} (\textbf{q})$ defined in the text, plotted as a function of phonon momentum $\mathbf{q}$. Results are shown for H-MLG (top) and \ce{CrI3} (bottom), for acoustic magnons with momenta ${\mathbf{Q}=\Gamma,K,M}$ in respective panels from left to right.} 
\end{figure}
Here, we compute the phonon dispersions and $e$-ph perturbation potentials with density functional perturbation theory~\cite{baroni_phonons_2001} in \textsc{Quantum ESPRESSO} \cite{Giannozzi2009}. We use the \textsc{Perturbo} code \cite{Zhou2021} to compute the $e$-ph interactions and combine them with BSE results to obtain the mag-ph couplings in Eq.~\eqref{eq:mag-ph}. This adds only a small overhead to the workflow as the computational cost is dominated by the BSE for magnon wave functions and DFPT for $e$-ph interactions.
\\
\indent
Physical intuition suggests that atomic vibrations coupling strongly to electrons may also couple strongly to their magnetic moment, but quantifying spin-phonon and mag-ph coupling is challenging~\cite{Jinsoo_PRL, Stengel_PRX}. Our method enables direct comparison of $e$-ph and mag-ph interactions, resolved for each phonon mode. Figure~\ref{fig:magph-mat}(b) shows the phonon dispersion with the strength of mag-ph and $e$-ph interactions color-coded in separate panels. The mag-ph is generally weaker than the $e$-ph interaction, and the strength of the two interactions exhibits distinct patterns across phonon modes. 
We find that specific $A_1$ phonon modes with high symmetry couple strongly to magnons, both in H-MLG and \ce{CrI3}. 
In H-MLG, the mode with the largest $e$-ph coupling is an out-of-plane C-H bond stretching mode, while the mode dominating mag-ph coupling at low energy is an out-of-plane ZO mode.

\indent
The differences between the $e$-ph and mag-ph coupling are apparent in \ce{CrI3} where, similar to H-MLG, the $e$-ph coupling is strongest for an in-plane LO mode, while the mag-ph interaction is dominated by an out-of-plane ZO mode with $\sim$18 meV energy. These strongly coupled atomic vibrations, illustrated in Fig.~\ref{fig:magph-mat}(c), are breathing modes of the iodine atoms surrounding the magnetic Cr ion which couple strongly to the magnetic moment but not to the electron charge. 
These results point to a strong coupling between out-of-plane modes and magnetic moments with layer-normal alignment in ferromagnetic monolayer CrI$_3$. 

\begin{figure}[t]
\includegraphics[width=\linewidth]{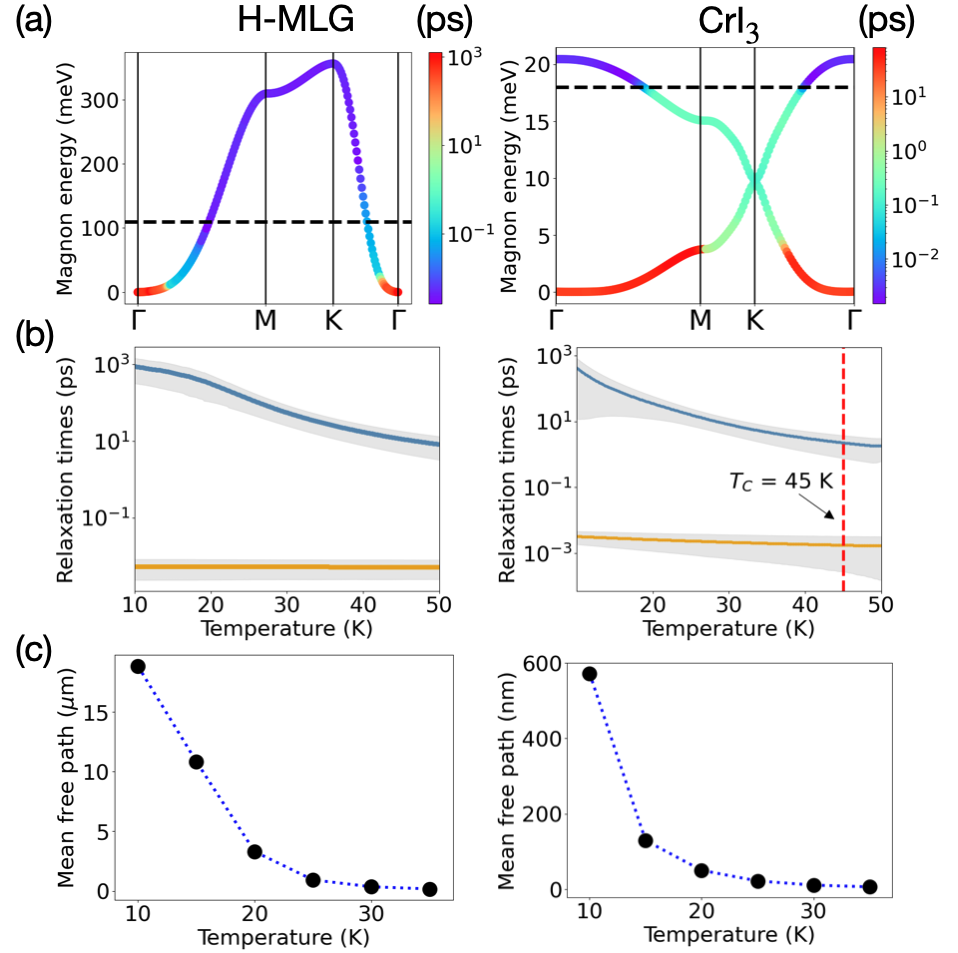}
\caption{\label{fig:tau} (a) Magnon dispersions with color-coded mag-ph relaxation times at 20~K. The dashed lines indicate the energy of the phonon with the strongest mag-ph couplin in each material. 
(b) Mag-ph relaxation times as a function of temperature. In each plot, the orange lines are averages over short-lived magnon states, while the blue lines are averages over long-lived magnon states, as explained in the text. (c) Average mean free paths $\overline{L}(T)$ in H-MLG (left) and \ce{CrI3} (right), obtained by taking a thermal average of the magnon mode-dependent mean free paths.}
\end{figure}

\indent
Analyzing the mag-ph interactions in momentum space can provide additional microscopic insight. We focus on the dependence of mag-ph scattering on transferred (phonon) momentum $\mathbf{q}$ for a fixed magnon momentum $\mathbf{Q}$. In Fig.~\ref{fig:magph-vs-q}, we plot mag-ph couplings averaged over all $N_{\rm m}$ magnon modes, $\overline{G}_{n\mathbf{Q}} (\mathbf{q}) = \sqrt{\sum_{m\nu} |G_{nm\nu} (\mathbf{Q},\mathbf{q})|^{2}/N_{\rm m}}$, as a function of phonon momentum $\mathbf{q}$ for selected magnon momenta. 
The average mag-ph couplings depend weakly on phonon momentum near the zone center (long-wavelength limit $\mathbf{q}\!\to \!0$) and vary more rapidly for larger phonon momenta near the Brillouin zone edge. The symmetry of the patterns decreases progressively as the magnon momentum $\mathbf{Q}$ goes from $\Gamma$ to $K$ to $M$. This symmetry reflects the little groups of the points $K$ and $M$ for the respective point group of the material, $C_{3v}$ in H-MLG and $D_{3d}$ in CrI$_3$.
\\
\indent
\textit{Magnon dynamics.\textemdash} Starting from the mag-ph interactions, we compute the corresponding relaxation time for each magnon mode $|\lambda_{n \mathbf{Q}}\rangle$ using lowest-order perturbation theory~\cite{SM}:
\begin{align}
\label{eq:gamma}
&(\tau^{\text{mag-ph}}_{n\mathbf{Q}}(T))^{-1} = \frac{2\pi}{\hbar}\frac{1}{\mathcal{N}_{\mathbf{q}}}
\sum_{m\nu q} |\mathcal{G}_{nm\nu}(\mathbf{Q}, \mathbf{q})|^{2} \nonumber \\
&\quad \times \big[ (N_{\nu \mathbf{q}} + 1 + F_{m\mathbf{Q} + \mathbf{q}}) \times \delta(E_{n\mathbf{Q}} - E_{m\mathbf{Q} + \mathbf{q}} - \hbar \omega_{\nu \mathbf{q}}) \nonumber \\
&\qquad+(N_{\nu \mathbf{q}} - F_{m\mathbf{Q} + \mathbf{q}}) \times \delta(E_{n\mathbf{Q}} - E_{m\mathbf{Q} + \mathbf{q}} + \hbar \omega_{\nu \mathbf{q}}) \big],
\end{align}
where $N(T)$ and $F(T)$ are phonon and magnon Bose-Einstein occupations at temperature $T$, respectively, and $\mathcal{N}_{\mathbf{q}}$ is the number of sampled $\mathbf{q}$-points. 
\\
\indent
We calculate the mag-ph relaxation times in H-MLG and CrI$_3$ at 20~K and map them on the magnon dispersions in Fig.~\ref{fig:tau}(a). At this relatively low temperature, which is below the Curie temperature of both materials, magnon scattering occurs primarily by phonon emission [$N_{\nu \mathbf{q}}(T)+1$ term in the second line of Eq.~\eqref{eq:gamma}] because the thermal phonon occupations are small, particularly for optical phonons. As a result, we find long relaxation times of about 1 ns in H-MLG and 10$-$100 ps in CrI$_3$ at low magnon energies in the long-wavelength limit $\mathbf{Q} \!\to \!0$. 
The relaxation times progressively decrease up to a certain energy in both materials. Above that threshold, which is shown with a dashed line for each material in Fig.~\ref{fig:tau}(a), emission of phonons with strong mag-ph coupling becomes energetically possible.  
\\
\indent
In H-MLG, the threshold occurs at $\sim$100 meV magnon energy, which corresponds to the energy of the flexural ZO mode with strong mag-ph coupling highlighted in Fig.~\ref{fig:magph-vs-q}(b). Above that threshold, emission of the strongly-coupled ZO phonon becomes possible, and the magnon relaxation times drop by orders of magnitude, reaching values of $\sim$10 fs near the highest magnon energy. 
This trend is in sharp contrast with $e$-ph interactions in both graphene and H-MLG, where flexural modes couple weakly to electronic charge; in graphene, this enhances the electrical and thermal conductivities to record large values~\cite{Mariani2008, Lindsay2010}. 
We observe similar trends in \ce{CrI3}, where a ZO optical phonon with 18 meV energy governs mag-ph interactions and defines a threshold for phonon emission, above which the magnon relaxation times drop by 2$-$3 orders of magnitude. (The rapid decrease in relaxation times at 5-meV magnon energy is due to relatively strong coupling with an acoustic phonon). Similar to H-MLG, the $e$-ph interactions per se are not particularly strong for the modes with the largest mag-ph coupling that govern magnon dynamics, highlighting key differences between the $e$-ph and mag-ph interactions. 
%
We have verified the validity of our pertubative approach by computing the ratio 
$R = \tau^{-1}_{nQ} / \omega_{nQ}$, where $R<1$ corresponds to well-defined magnon quasiparticles. We find $R<1$ over most of the magnon spectrum in both materials, except for magnons nearly degenerate to strongly-coupled phonons \cite{SM}.  
Our calculations correctly predict that those magnons will be short-lived ($\sim$10 fs) due to strong mag-ph coupling.
\\
\indent
The presence of strongly coupled phonons causes the mag-ph relaxation time distribution to become nearly bimodal, separating short-lived and long-lived magnons with energies, respectively, above and below the emission threshold. 
In Fig.~\ref{fig:tau}(b), we plot the average relaxation times as a function of temperature up to 50~K for these two groups of magnons. (Note that this analysis is meaningful only below the Curie temperature, which is 45~K in CrI$_3$ and much higher in H-MLG).
The order-of-magnitude relaxation time difference between short- and long-lived magnons implies that nonequilibrium magnons with a broad energy distribution would rapidly relax to magnons with subthreshold energy, and thus key transport phenomena relevant in magnon devices will occur primarily in the low-energy, long-lived magnon regime.
\\
\indent
Therefore, we compute the average mag-ph mean free paths (MFPs) in this subthreshold window by combining first-principles magnon velocities (obtained from magnon dispersions) and relaxation times. We first compute the MFP for each magnon mode, $L_{n\textbf{Q}}(T)=v_{n\textbf{Q}}\tau_{n\textbf{Q}}(T)$, where $v_{n\textbf{Q}}$ is the magnon group velocity and $\tau_{n\textbf{Q}}(T)$ the mag-ph relaxation time, 
and then take a thermal average using $\overline{L}(T)= \sum_{n\mathbf{Q}} L_{n\mathbf{Q}}(T) F_{n\mathbf{Q}}(T)$, where the sum extends over the subthreshold window and $F$ is the magnon thermal occupation.
Figure~\ref{fig:tau}(c) shows the average magnon MFPs computed as a function of temperature. The MFPs in H-MLG are approximately 10 $\mu$m at 15 K and decrease rapidly to 0.1 $\mu$m at \mbox{40~K,} while in \ce{CrI3} the average MFP is 0.13 $\mu$m at 15~K and decreases to 4 nm at 40~K. The rapid decrease of the mean free path with temperature is due to scattering with low-energy phonons, whose coupling with magnons is non-negligible [see Fig.~\ref{fig:magph-mat}(b)], as well as increased scattering phase space for thermally excited magnons.
Interestingly, the MFP values in H-MLG are comparable to those in yttrium iron garnet (YIG), a material studied extensively for its unique magnon physics, where measurements have shown magnon MFPs of $\sim$10~$\mu$m in the same temperature range~\cite{Boona2014_2}. Such quantitative estimates of magnon MFPs can guide the development of magnon-based devices. 

\indent
\textit{Discussion.\textemdash} In this Letter, we have shown a first-principles approach to study mag-ph interactions in real materials. 
Our method repurposes the \textit{ab initio} BSE, a well-established technique for excitons, to study magnon dispersions and mag-ph interactions, allowing parameter-free calculations of magnon relaxation times and mean free paths. Our approach can describe how magnons interact with specific phonon modes depending on phonon symmetry and character, features that cannot be easily resolved in phenomenological spin models.
This advance provides a quantitative tool to study magnon physics, including magnon transport and coupled magnon and phonon phenomena, with applications to broad classes of magnetic materials and magnon-based devices. 
Focusing on CrI$_3$ as a case study, we have predicted that in-plane atomic vibrations couple to charge, while out-of-plane atomic vibrations couple to magnetic moments. This insight suggests opportunities for controlling charge and spin transport in Cr trihalides and may extend to other 2D ferromagnets. 
\\

\indent
Our approach can be extended to antiferromagnetic and noncollinear spin systems, and materials with SOC. In particular, if SOC is included in the BSE and $e$-ph interactions~\cite{Marsili2021,Zhou2021}, spin-flip $e$-ph scattering becomes possible, leading to mixing and phonon-mediated coupling of excitons and magnons. We also neglect higher-order magnon-two-phonon interactions $b^\dag b(S+S^\dag)$.
Future work will explore these directions, as well as extend the calculations to bulk materials and real-time coupled magnon and phonon dynamics, mirroring progress in exciton-phonon calculations~\cite{Chen2022}. This may enable the exploration of intriguing phenomena such as magnon-polaron effects, and phonon-driven magnetism, among others.\\

K.B.L. and M.B. were supported by the National Science Foundation under Grant No. OAC-2209262 for code and method development, and by the AFOSR and Clarkson Aerospace under Grants No. FA9550-21-1-0460 and FA9550-24-1-0004 for calculations on 2D materials. 
This work is supported by the European Union’s Horizon Europe research and innovation program under the Marie Sklodowska-Curie grant agreement 101118915 (TIMES). 
This work is part of the project I+D+i PID2023-146181OB-I00 UTOPIA, funded by MCIN/AEI/10.13039/501100011033, project PROMETEO/2021/082 (ENIGMA) and SEJIGENT/2021/034 (2D-MAGNONICS) funded by the Generalitat Valenciana. 
This study is also part of the Advanced Materials program (project SPINO2D), supported by MCIN with funding from European Union NextGenerationEU (PRTR-C17.I1) and by Generalitat Valenciana. 
A. M.-S. acknowledges the Ramon y Cajal program (grant RYC2018-024024-I; MINECO, Spain). 
A.E.-K. acknowledges the Contrato Predoctoral Ref. PRE2021-097581. D.S. acknowledges funding from MaX ”MAterials design at the eXascale” (Grant Agreement No. 101093374) cofunded by the European High Performance Computing joint Undertaking (JU) and participating countries, and from the Innovation Study Isolv-BSE that has received funding through the Inno4scale project, which is funded by the European High-Performance Computing Joint Undertaking (JU) (Grant Agreement No 101118139).\\

\textit{Note added.\textemdash}While submitting this manuscript, we became aware of a recent paper with a similar title \cite{Fang2025}, which focuses on a different problem, mag-ph hybridization using density functional theory. 

\providecommand{\noopsort}[1]{}\providecommand{\singleletter}[1]{#1}%

\end{document}